\documentclass[
journal=jacsat, %
manuscript=article]{achemso}

\usepackage{subfigure}
\usepackage{color}

\usepackage{url}
\usepackage{hyperref}

\author{J. M. de Sousa$^{1,2,3,*}$, A. L. Aguiar$^2$, E. C. Gir\~ao$^2$, A. F. Fonseca$^{1}$, A. G. Souza Filho$^{3}$, and Douglas S. Galvao$^{1,}$}

\affiliation{$^1$Instituto de F\'isica Gleb Wataghin, Universidade Estadual de Campinas, 13083-970, Campinas, SP, Brazil.}
\affiliation{$^2$Departamento de F\'isica, Universidade Federal do Piau\'i, Teresina, Piau\'i, 64049-550, Brazil}
\affiliation{$^3$Departamento de F\'isica, Universidade Federal do Cear\'a, P.O. Box 6030, CEP 60455-900, Fortaleza, Cear\'a, Brazil.}
\email{galvao@ifi.unicamp.br  ; josemoreiradesousa@gmail.com}

\title[\texttt{achemso} demonstration]
{Mechanical Properties and Fracture Patterns of Pentagraphene Membranes}
\begin{document}
\begin{abstract}
  Recently, a new two-dimensional carbon allotrope called
  pentagraphene (PG) was proposed. PG exhibits mechanical and electronic interesting properties, including typical band gap values of semiconducting materials. PG has a Cairo-tiling-like 2D lattice
  of non coplanar pentagons and its mechanical properties 
  have not been yet fully investigated. In this
  work, we combined density functional theory (DFT) calculations and
  reactive molecular dynamics (MD) simulations to investigate the mechanical properties and fracture patterns of PG membranes under tensile strain. We show that PG membranes can hold up to 20\% of strain and that fracture occurs only after substantial dynamical bond breaking and the formation of 7, 8 and 11 carbon rings and carbon chains. The stress-strain behavior was observed to follow two regimes, one exhibiting linear elasticity followed by a plastic one, involving carbon atom re-hybridization with the formation of carbon rings and chains.
 Our results also show that mechanically induced structural transitions from PG to graphene is unlikely to occur, in contrast to what was previously speculated in the literature.


\end{abstract}



\section{Introduction}

\hspace*{0.50cm} Graphene is one of the most important topics in materials science today ~\cite{novoselov2005two,ruoff2009,aja2009,stankovich2006graphene,geim2009graphene,chen2008mechanically,eda2009graphene,scarpa2009effective}.
Several studies have focused on physical and/or chemical
modifications of the perfect honeycomb lattice, since its zero band gap value limits the development of some
pure graphene-based digital electronic
devices~\cite{withers2010electron}. Functionalization of graphene ~\cite{xu2008flexible,ramanathan2008functionalized}
and graphene nanoribbons~\cite{son2006half} are examples of strategies
used to tune the band gap, but which has achieved only partial success. Due to of this, there is a renewed interest on other layered
structures which have a band gap. Hexagonal
boron-nitride~\cite{giovannetti2007substrate,watanabe2004direct},
carbon nitride
nanosheets~\cite{niu2012graphene,thomas2008graphitic,moreiracn}, metal
dichalcogenides~\cite{wang2012electronics,chhowalla2013chemistry}, and
silicene
membranes~\cite{takeda1994theoretical,aufray2010graphene,vogt2012silicene}
are examples of other two-dimensional structures that overcome
graphene ``{\it bandgapless}'' limitation. Other pure carbon structures such as
graphynes~\cite{baughman1987structure,coluci2003families,coluci2004new,reviewgraphyne1,reviewgraphyne2} and haeckelites~\cite{terrones2000} are also good candidates.

\begin{figure}[htb!]
 \centering
 \includegraphics[scale=0.4]{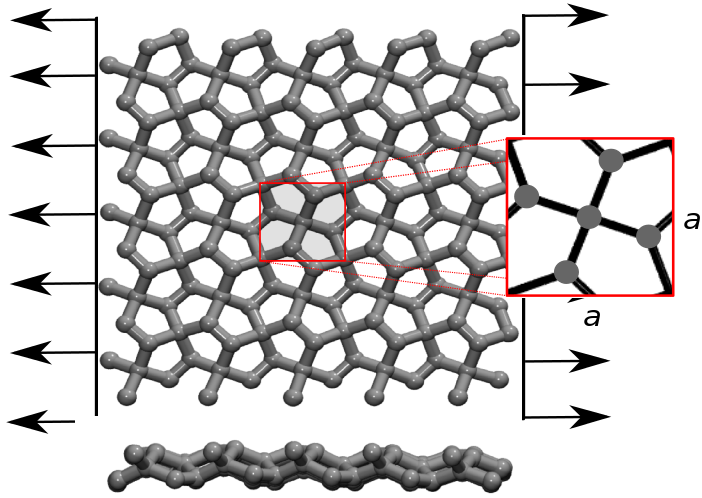}
 \caption{Frontal (top) and lateral (bottom) views of a pentagraphene (PG) membrane. The
   arrows indicate the directions of the applied tensile deformations
   considered in this study. The figure inset shows the PG square unit cell, with lattice parameter a = 3,64~\AA.}
\label{fig1}
\end{figure}

Recently, a new 2D carbon allotrope called {\it pentagraphene} (PG) (see Figure 1) has
been proposed by Zhang {\it et al}.~\cite{zhang2015penta}. Based on
DFT calculations, they showed that such membranes have a unique
arrangement of carbon atoms in a network of non-coplanar pentagons,
similar to a {\it Cairo pentagonal tiling}. They also showed that
PG is not only mechanically and thermodynamically stable,
but also presents a large band gap of $3.25$eV ~\cite{zhang2015penta}. Besides that, PG also exhibits interesting thermal and mechanical properties, such as
negative Poisson's ratio (auxetic behaviour ~\cite{baughman1993}) due to its metastability and
intricate atomic structural configuration. Figure~\ref{fig1} shows PG frontal and a side
view.

Several theoretical studies have been already devoted to PG and its stability has been the subject of debate ~\cite{questionpenta}. Some first-principles studies suggest PG is stable and described some of their electronic and mechanical
properties~\cite{dftpentajap2015,dftpentascirep2016}. PG thermal
conductivity at room temperature has been estimated
by different methods to be about 167 W/mK~\cite{pentaTC1} (from MD
simulations) and 645 W/mK~\cite{pentaTC2} (from first principles
calculations). PG nanoribbons have also been theoretically investigated in terms of stability and electronic band structure. These quasi-1D systems have been
shown to preserve the semiconducting character from their layer
parent counterpart and the gap value depends on width ~\cite{pentaribbon}. Similar to graphene, PG functionalization
(hydrogenation and fluorination) allows the tuning of its electronic and
mechanical properties~\cite{liTunningPenta}, and a unexpected
increase in thermal conductivity was observed in the
hydrogenation case~\cite{TCincreaseHpenta}. Recently, interesting
PG mechanical and structural behaviors  were reported
based on reactive molecular dynamics (MD)
simulations~\cite{cranford6to5}. It has been predicted that a structural transition
from PG to graphene (or from PG to hexagraphene) can occur as a result of thermal and/or tensile
strains, thus leading to the assumption that PG and graphene might be
considered different structural phases of the same material~\cite{cranford6to5}.


In this paper, we combine density functional theory (DFT) calculations and
reactive classical MD simulations (with a properly chosen set of
potential parameters) to investigate the fracture patterns of PG membranes under axial tensile strain, as schematically shown in Fig.~\ref{fig1}. Differently from the conclusions of Ref.~\cite{cranford6to5}, our results
reveal the formation of structures with 7, 8 and 11 carbon rings and carbon chains, just before the mechanical failure (fracture),which happens at
about 20\% of strain. Both Young's modulus and Poisson's ratio of PG were also calculated and compared with the original {\it ab initio} predictions~\cite{zhang2015penta}.

\section{Methods}

\hspace*{0.50cm}We combined quantum (DFT) and classical (MD) methods to investigate
the structural and dynamical aspects of PG membranes under tensile strain, up to the limit of mechanical failure (fracture). 
In the following sections we provide technical
details on the computational techniques used in this paper.

\subsection{DFT calculations}

\hspace*{0.50cm}We used a LCAO-based DFT approach~\cite{hohenberg64,Kohn65}, as
implemented in the SIESTA code~\cite{ordejon96,portal97}. The
Kohn-Sham orbitals were expanded in a double-$\zeta$ basis set
composed of numerical pseudoatomic orbitals of finite range enhanced
with polarization orbitals. A common atomic confinement determined by
an energy shift of $0.02$ $Ry$ was used to define the cutoff radius
for the basis functions, while the fineness of the real space grid was
determined by a mesh cutoff of 400 Ry~\cite{anglada02}. For the
exchange-correlation potential, we used the generalized gradient
approximation (GGA)~\cite{perdew96}, and the pseudopotentials were
modeled within the norm-conserving Troullier-Martins
\cite{troullier91} scheme in the Kleinman-Bylander \cite{kleinman82}
factorized form. Brillouin-zone integrations were performed by using a
Monkhorst-Pack\cite{monkhorst76} grid of $8$ $\times$ $8$ $\times$ $1$
$k$-points. All geometries were fully optimized for each strain level
until the maximum force component on any atom was less than 10
meV/\AA.  
The lattice vectors were manually deformed along selected directions
(uniaxial and biaxial) and the coordinates of carbon atoms were
rescaled along these directions before fully convergence. For uniaxial
stretching, we have considered two cases: with and without
constrains along the perpendicular directions. The stress tensor
$\sigma_{ij}$ is related to strain tensor $\varepsilon_{ij}$
($i,j=x,y,z$) by $\sigma_{ij}=(1/S)(\partial
U/\partial\varepsilon_{ij})$, where $S=(\vec{a_x}\times\vec{a_y})$ is
the area of the unit cell.
For each strained structural geometry relaxation, the SCF
convergence thresholds for electronic total energy were set to
$10^{-4}$ eV.  Periodic boundary conditions were imposed, with a
perpendicular off-plane lattice vector a$_z$ large enough ($20$ \AA) to prevent spurious interactions between periodic images.

\subsection{MD simulations}

\hspace*{0.50cm}The MD simulations were performed using the reactive force field
(ReaxFF)~\cite{van2001reaxff,mueller2010development}. The numerical
integration of the Newton's equations was performed in the large-scale
atomic / molecular massively parallel simulator (LAMMPS)
code~\cite{plimpton1995fast}. ReaxFF is a reactive force field
developed by van Duin, Goddard III and co-workers, which is designed
to be a bridge between quantum chemical force fields and empirical
bonding energy terms. ReaxFF is parameterized using available experimental data and/or using DFT calculations. In ReaxFF, the total bond
energy between atoms are obtained through the computation of all
interatomic distances and updated at every time step of the classical
MD runs.  In this way, the structural connectivity is determined uniquely
by the atomic positions, thus allowing the ReaxFF to create and
break (dissociate) chemical bonds in a dynamical way, through the whole simulation. This is important to describe not only the equilibrium structures, but also the fracture patterns of the investigated systems. The energy of the system is divided into partial energy contributions,
which include bonded and non-bonded terms as follows ~\cite{van2001reaxff}:
\begin{eqnarray}
\label{esys}
E_{system}&=&E_{bond}+E_{over}+E_{under}+E_{val}\nonumber \\
&&+E_{pen}+E_{tor}+E_{conj}+E_{vdW}\nonumber\\
&&+E_{co} \quad , 
\end{eqnarray}
where each term, respectively, represents the energies corresponding to
the bond distance, the over-coordination, the under-coordination, the
valence, the penalty for handling atoms with two double bonds, the
torsion, the conjugated bond energies, the van der Waals, and coulomb
interactions, respectively.

ReaxFF has been extensively used in the study of the dynamic aspects
of nanostructures, such as fractures of
graphynes~\cite{cranford2011mechanical}, silicene membranes
~\cite{botari2014mechanical}, connected carbon
nanorings~\cite{chen2005mechanical},
carbyne~\cite{mirzaeifar2014tensile,nair2011minimal}, the degradation
of graphene and graphdiyne membranes in gaseous
atmospheres~\cite{flores2009graphene,autreto2014site}, among other carbon based nanostructures.

Here, for the study of structural and fracture mechanics of
PG structures, we considered square membranes under periodic boundary conditions
with dimensions of approximately 80 \AA $\times$ 80 \AA. In all calculations, these structures were initially thermalized at 300
K in a NPT ensemble, in order to obtain a structure corresponding to
zero external pressure, before the beginning of the fracture dynamics
study. After that, a stretching process was then considered within a NVT
ensemble, also at 300 K, with the temperature set controlled by a Nose-Hoover
thermostat~\cite{evans1985nose}, as implemented in the LAMMPS~\cite{plimpton1995fast} code.

In our calculations, the timestep of numerical integration was set to
$0.05$ fs and a constant strain rate of $10^{-6}$ fs$^{-1}$ was
considered. The above conditions were maintained up to the mechanical failure limit. The PG mechanical properties were analyzed by the stress-strain relationship, where
the engineering strain, $\varepsilon$, under tension is defined as
\begin{eqnarray}
\label{epis}
\centering
\varepsilon = \frac{\zeta - \zeta _{0}}{\zeta _{0}}= \frac{\Delta \zeta }{\zeta _{0}},
\end{eqnarray}
where $\zeta_0$ and $\zeta$ are the length of the structure before and
after the dynamics of deformation, respectively. The {\it per-atom}
stress tensor of each carbon atom are calculated
by~\cite{garcia2010bioinspired}:
\begin{eqnarray}
  \sigma_{\alpha \beta }=\frac{1}{\Gamma }\sum_{i}^{N}\left (m_{i}v_{\alpha i}v_{i\beta } + r_{i\alpha }f_{i\beta } \right ) , 
\end{eqnarray}
where 
$\Gamma$ is the atom volume, $N$ the number of atoms, $m_i$ the mass
of carbon atoms, $v$ the velocity, $r$ the coordinates of the carbon
atoms and $f_{i\beta}$ is the $\beta$ component of the force acting on
the $i$-{\it th} atom. In order to perform a more detailed analysis of the distribution of
stress along the structure during the fracture process, we
also calculated the quantity known as {\it von Mises stress},
$\sigma_{vM}$, which is mathematically given
by~\cite{garcia2010bioinspired}:
\begin{eqnarray}
  \sigma_{vM}=\left [ \frac{(\sigma_{xx}-\sigma_{yy})^{2}+
      (\sigma_{yy}-\sigma_{zz})^{2}+(\sigma_{zz}-\sigma_{xx})^{2}+
      (\sigma_{xy}+\sigma_{yz}+\sigma_{zx})^{2}}{2} \right ]^{\frac{1}{2}} , 
\end{eqnarray}
where $\sigma_{xy}$, $\sigma_{yz}$ and $\sigma_{zx}$ are shear
stress components. The von Mises stress has been used in the mechanical studies
of other nanostructures as silicene
membranes~\cite{botari2014mechanical} and carbon
nanotubes unzipping~\cite{dos2012unzipping}. It is very useful to visualize how the stress accumulates and dissipates during the stretching/fracture processes.

\section{Results}

\subsection{Choice of ReaxFF parameters}

\hspace*{0.50cm}Before starting the MD study of PG fractures, we performed
DFT- and MD-based calculations tests to use as benchmark for the choice of the multiple avaialble ReaxFF set parameters. Among the possible choices, we considered four different ReaxFF sets, as developed by Mueller {\it et
  al.}~\cite{mueller2010development}, Mattsson {\it et
  al.}~\cite{mattsson2010first}, Chenoweth {\it et
  al.}~\cite{chenoCHO}, and Srinivasan {\it et
  al.}~\cite{goverapet2015development}. These parameters were
developed for carbon in different multicomponent systems. The one from
Srinivasan, for instance, was recently developed for condensed phases
of carbon. The tests consist in the calculation of the thickness, the
Young's modulus, $Y$, and Poisson's ratio, $\nu$, of
PG structures using the same protocols by Zhang {\it
  et al}.~\cite{zhang2015penta} to estimate the elastic constants
$C_{11}$ and $C_{12}$, and also using the following equations:
\begin{equation}
\label{youngs}
Y = \frac{C^{2}_{11}-C^{2}_{12}} {C_{11}} \, , 
\end{equation}
and 
\begin{equation}
\label{poisson}
\nu=\frac{C_{12}}{C_{11}} \, .
\end{equation}
The elastic constants $C_{11}$ and $C_{12}$ were obtained from energy
minimization calculations of the structure in uniaxial and biaxial
tensile strains, respectively. Uniaxial simulations, (as shown in
Fig.~\ref{fig1}), are made by fixing one dimension and applying
strain along the other direction. Biaxial tensile strain consists of
applying the same amount of strain along $x$ and $y$ directions at the
same time. The energy minimizations were calculated with convergence
tolerances of $0$ and $10^{-8}$ for the energy and force,
respectively.

In Table (\ref{tab}) we present the results previously reported in Ref.~\cite{zhang2015penta} for $Y$ and $\nu$, as well as the DFT and ReaxFF (four different set parameters) results obtained in our simulations.
\begin{table*}[ht] 
\centering
\caption{Comparison of structural and mechanical properties
  of pentagraphene (PG) structures obtained from DFT~\cite{zhang2015penta}, our DFT
  calculations, and ReaxFF
  \cite{mueller2010development,mattsson2010first,chenoCHO,goverapet2015development}.}
\begin{tabular}{cccc} 
\hline 
\hline 
Method & Thickness & Young's Modulus  & Poisson's ratio  \\ 
       &    (\AA)     &  (GPa.nm) &                        \\ 
\hline
\hline
DFT from Ref.~\cite{zhang2015penta} & 1.20 & 263.8 & -0.068 \\
DFT from our calculations & 1.23  & 257.6 & -0.096 \\ 
ReaxFF - Mattsson~\cite{mattsson2010first} & 0.882 & 150.5 & -0.154 \\
ReaxFF - Srinivasan~\cite{goverapet2015development} & 1.34 & 133.9 & 0.366 \\
ReaxFF - Muller~\cite{mueller2010development} & 1.05 & 322.0 & 0.335 \\
ReaxFF - Chenoweth~\cite{chenoCHO} & 1.09 & 197.0 & 0.380 \\
\hline
\hline
\end{tabular}
\label{tab}
\end{table*} 
From this table we can see a good agreement between our DFT (performed with a
localized orbital basis) results and those from
Ref.~\cite{zhang2015penta} (which used a plane-wave basis set), including the prediction of the auxetic behavior. We then tested the different ReaxFF sets in order to determine which one provides the best results in comparison to the DFT ones, in terms of structural and mechanical properties. We observe
that the Mattsson~\cite{mattsson2010first} set of parameters is the
only one which correctly predicts the sign of PG Poisson's
ratio. However, it presents a much softer structure (smaller thickness
and modulus than those calculated from DFT). The
Srinivasan~\cite{goverapet2015development} set of parameters predicts
reasonable thickness but a much smaller elastic modulus than that from
DFT. On the other hand, the Mueller~\cite{mueller2010development} and
the Chenoweth~\cite{chenoCHO} sets of parameters present the best
matches for the thickness and Young's modulus as compared with both
DFT calculations. If we take the results from Zhang {\it et
  al.}~\cite{zhang2015penta} as reference, both Mueller's and the
Chenoweth's sets give smaller thickness with similar deviations. On
the other hand, those two parameter sets show distinct trends for the
Young's Modulus: $Y$ is $\sim22\%$ higher for Muller's and $\sim25\%$
lower for Chenoweth's in comparison with Zhang's DFT result. For the physical phenomenon aspects we are investigating here, the thickness and Young's modulus parameters are the most important to be precisely described. In this sense the Chenoweth's ~\cite{chenoCHO} and Mueller's \cite{mueller2010development} set of parameters would be the best choice.
As Chenoweth~\cite{chenoCHO} set was already used in
Ref.~\cite{cranford6to5} to investigate PG tensile strain tests, we decided to use Mueller's
set~\cite{mueller2010development} in our calculations, so we would have a good reference for comparisons.

\subsection{MD results}

\begin{figure}[htb!]
 \centering
 \includegraphics[scale=0.5]{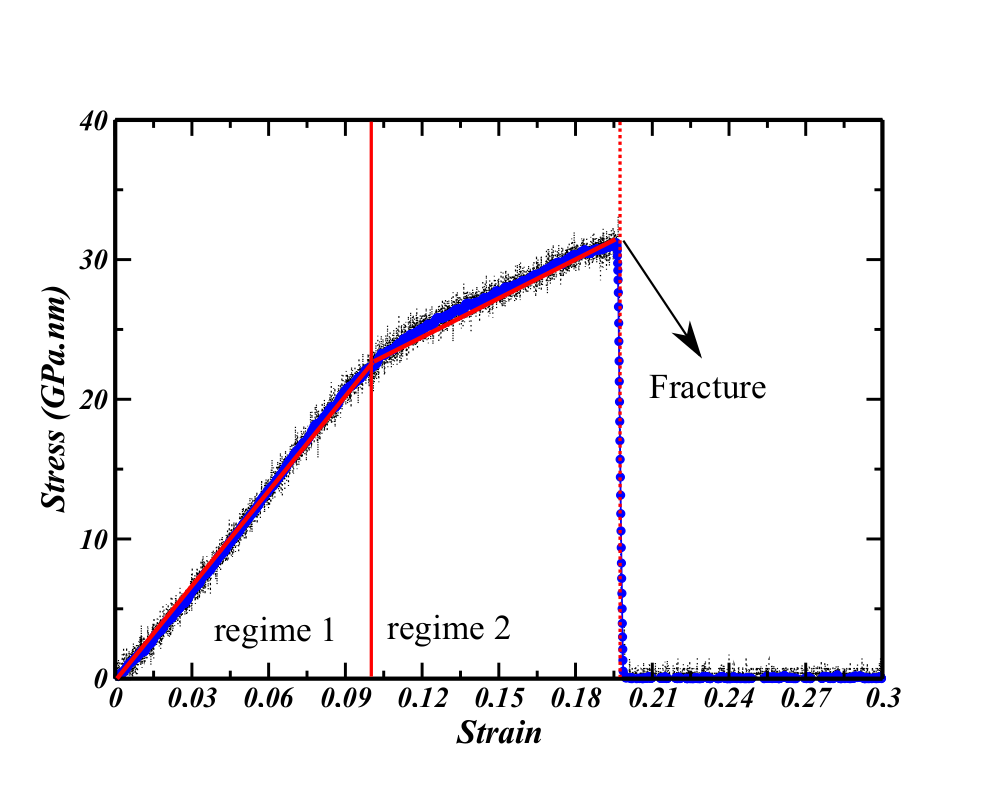}
 \caption{Pentagraphene (PG) stress-strain curve.
   Vertical red lines divide the curve into elastic (regime 1) and plastic (regime 2). Over-imposed red line on the stress-strain curve indicates the average inclinations in both regimes. Dashed vertical line indicates the maximum strain the system can stand before fracture.}
\label{stst}
\end{figure}

\begin{figure}[htb!]
 \centering
 \includegraphics[scale=0.37]{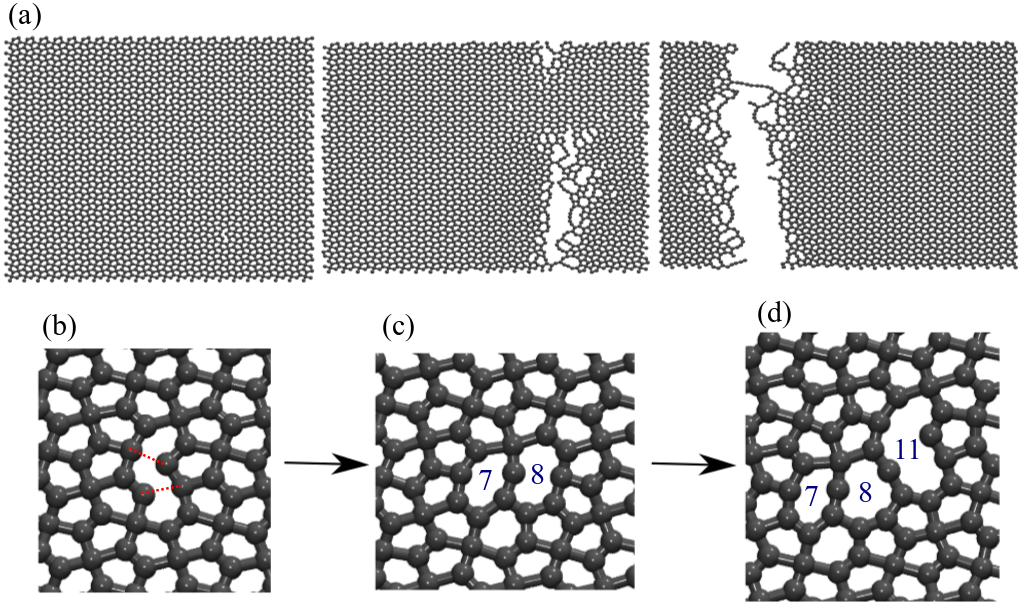}
 \caption{(a) Representative MD snapshots of tensioned PG structures at 0\% (leftmost panel), 19.7\% (middle panel) and 20\% (rightmost panel) strains. (b), (c) and (d) are representative MD snapshots showing the existence of 7, 8, 11 rings formed at 18\%, 18.5\% and 18.7\% tension strains.}
\label{fracture}
\end{figure}

\hspace*{0.50cm}In Figure~\ref{stst} we present the PG stress-strain curve obtained from
classical MD simulations based on the
Mueller~\cite{mueller2010development} set of ReaxFF parameters. We
observe two linear regimes, one elastic (regime 1) and the other a plastic one (regime 2) resulting from permanent deformations due to local structural reconstructions, as shown in
Fig.~\ref{fracture}. While the plastic regime starts at
about 10\% of strain, fracture takes place at about 20\%, which is very close to the maximum of 21\% of bi-axial tensile
strain calculated by DFT~\cite{zhang2015penta}.

\begin{figure}[htb!]
 \centering
 \includegraphics[scale=0.37]{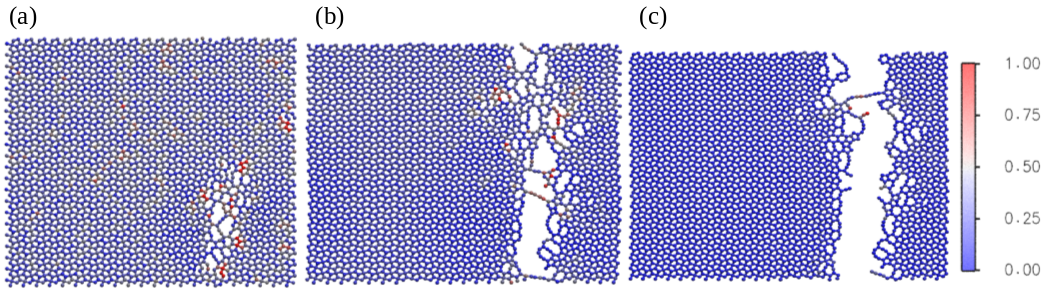}
 \caption{Representative MD snapshots showing the von Mises stress values of PG tensioned structures at (a) 19\%, (b) 19.5\% and (c) 20\% strains.}
\label{vonmises}
\end{figure}

\begin{figure}[htb!]
 \centering
 \includegraphics[scale=0.42]{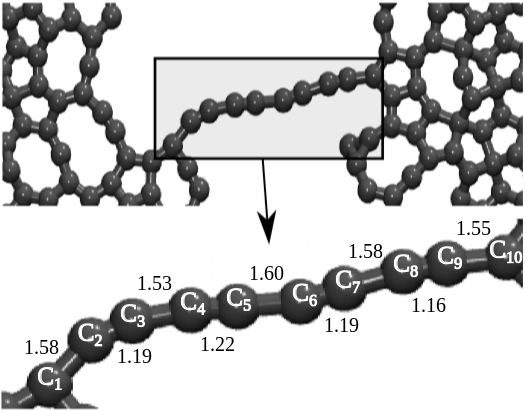}
 \caption{Structural details of a carbon chain formed at the last stages (before breaking) of a PG tensioned structure. Carbon
   atoms are labelled to identify the single and triple bonds that
   forms along the chain.Values in Angstroms.} 
 \label{chains}
\end{figure}

In Figure~\ref{fracture}a we present MD snapshots of PG strained
structures, including one close to the moment of fracture, where the formation of many carbon chains can be seen. In Figures~\ref{fracture}b-d we present representative MD snapshots of the plastic regime, where it is possible to observe the existence of 7, 8 and 11 carbon rings. These rings are formed from the reconstruction of broken $C-C$ bonds. In figure~\ref{vonmises}, we present representative MD snapshots showing the von Mises stress values of PG tensioned structures. Details of
the carbon chains that are formed at large tension strains, just before
final rupture of the structure, are shown in Fig.~\ref{chains}. The distances between the carbon atoms along the chains
indicate the formation of a structure having single and triple bonds which is the so called polyynic configuration. This chain configuration has been predicted to be the most stable linear structure.

These results are quite different from those reported in
Ref.~\cite{cranford6to5}, which predicted that the PG structures
evolve mostly to graphene-like conformations, during either
tensile or thermal strains, although structural defects were also present. In our simulations, we did not observe the formation of hexagons during the tensile strains. The difference between our calculations and those reported in
Ref.~\cite{cranford6to5} is mainly the choice of the set of ReaxFF
parameters. Ref.~\cite{cranford6to5} used the
Chenoweth~\cite{chenoCHO} set of parameters while we used the
Mueller~\cite{mueller2010development} one. From Table (\ref{tab}),
and as discussed before, we see that the main differences between them
are the results for the PG elastic modulus values. Chenoweth~\cite{chenoCHO}/(Mueller~\cite{mueller2010development}) parameters provide softer (harder) PG structures than that predicted by DFT. For this reason, the harder structure predicted by
Muller's set can transform itself from pentagons directly to higher order
rings, rather than to intermediate hexagons. This different results point out to the importance of a good choice of parameters in simulating the systems in order to prevent predictions that come out from the unprecise description of the structure rather than from the physical behavior. 

In order to gain further insights on which results are more realistic, we performed DFT calculations of the tensile strain up to fracture, as discussed next.


\subsection{DFT results}

\begin{figure}[htb!]
 \centering
 \includegraphics[scale=0.185]{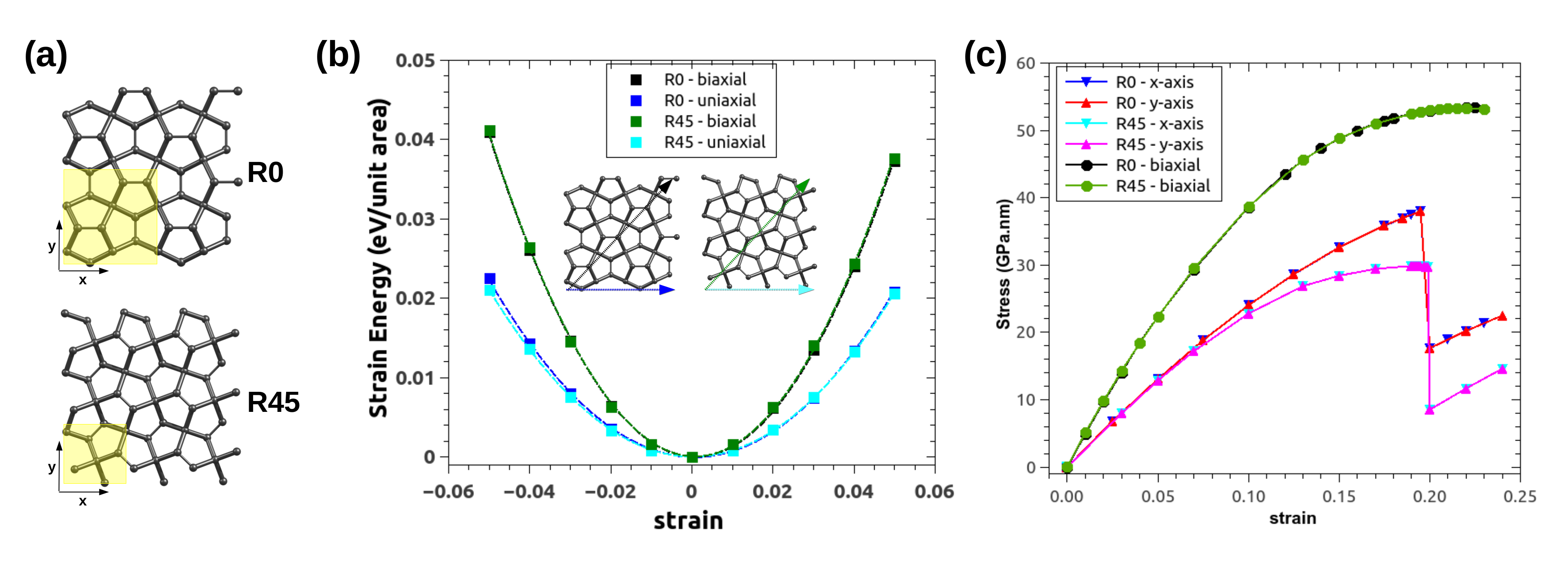}
 \caption{(a) R0 and R45 unit cells used to study pentagraphene (PG)
   failure mechanisms. The PG primitive cell (yellow
   square) was $2 \times 2$ ($3 \times 3$) replicated 
   for R0 (R45) structure in order to model the
   fracture process. (b) PG strain-energy curves in low-strain
   regime. Parabolic fitting was used to estimate the Young's
   modulus value (see text). (c) PG strain-stress curves for
   R0 and R45 under equi-biaxial and uniaxial
   stretching. Equi-biaxial (black and green) curves is
   ($\sigma_{xx}+\sigma_{yy}$) stress versus strain
   ($\varepsilon_{xx}$). Uniaxial curves is ($\sigma_{xx}) \sigma_{yy}$
   versus ($\varepsilon_{xx}$) $\varepsilon_{yy}$ strain.}
 \label{fig06}
\end{figure}

\hspace*{0.50cm}We have carried out DFT calculations of PG structures under uniaxial
and biaxial tensile strains up to limit of rupture. One of the goals of these calculations is to identify intermediate structures to help to understand the fracture dynamics. In addition, we want to get further insights on how suitable is the chosen set of ReaxFF parameters to describe the PG mechanical properties.
Two orthogonal PG units cells were used in order 
to investigate differences in failure mechanisms. These
units cells were labeled R0 and R45 (rotated of $45^\circ$ from R0)
as shown in Fig.~\ref{fig06}a. It is interesting
to note that R45 have neither perpendicular nor parallel C-C bonds along
uniaxial \textit{x} and \textit{y} strains. However, R0 structure has both types
of bonds, which will play a significant hole just before the rupture
under uniaxial strain.
We have calculated the energy shift due to the in-plane strain to
determine the PG mechanical stability. 
For a 2D
membrane, using the standard Voigt notation (1-\textit{xx}, 2-\textit{yy}, and 6-\textit{xy}),
the elastic strain energy per unit area can be expressed as a function
of C$_{11}$ , C$_{22}$ and C$_{12}$ elastic modulus tensor,
corresponding to second partial derivative of strain energy with
respect to strain. The elastic constants can be derived by fitting the
energy curves associated with uniaxial and equi-biaxial strains. The
curves are plotted in Fig.~\ref{fig06}b. 
We should note here that the mechanical behavior of R0 and R45 structures are
almost the same in low-strain regime (up to 5\%) producing similar results for the elastic constants. 
Under uniaxial strain,
$\varepsilon_{yy}=0$,
$U(\varepsilon_{xx})=1/2C_{11}\varepsilon^2_{xx}$. Parabolic fitting
of the uniaxial strain curve yields C$_{11}$ = 277.5
GPa$\cdot$nm. Under equi-biaxial strain,
$\varepsilon_{yy}=\varepsilon_{xx}$, we have
$U(\varepsilon_{xx})=(C_{11}+C_{12})\varepsilon^2_{xx}$. By
fitting the equi-biaxial strain curve we obtain $C_{11}+C_{12}$ =
250.8 GPa$\cdot$nm, hence, C$_{12}$ = -26.7 GPa$\cdot$nm.
The in-plane Young's modulus is calculated to be as large as
274.95 GPa$\cdot$nm, which is very similar to what was observed by other
authors~\cite{zhang2015penta}.
We also note that C$_{12}$ is negative
for this membrane, leading to a negative Poisson's ratio (NPR),
$\nu=C_{12}/C_{11}=-0.096$. This result confirms that PG is an
auxetic material.  We also studied the ideal PG strength and failure
mechanism by calculating the variation of stress as
a function of the equi-biaxial and uniaxial tensile strain. The results
are presented in Fig.~\ref{fig06}c, which shows that the strain at the
maximum stress before failure is 19.5\% (uniaxial) and 23\% (biaxial).

The simulation of uniaxial stretching loading with the in-plane perpendicular lattice vector fixed also allowed the computation of the residual perpendicular stress components (not shown in Fig.~\ref{fig06}c),
which exhibit negative values. Furthermore, 
when no constrains are imposed to uniaxial loading, the length of 
perpendicular lattice vectors increases. This result is an 
additional evidence that PG is an auxetic material. 
For equi-biaxial stretching loading, we plot ($\sigma_{xx}+\sigma_{yy}$) in
Fig~\ref{fig06}c. Therefore, we observed that the calculated ultimate
tensile strength (UTS) shows that PG is very strong with the UTS
of $\sim$38GPa.nm (R0 uniaxial), and $\sim$29GPa.nm (R45 uniaxial). This
discrepancy will be further discussed. For biaxial stretching, we
obtained UTS of $\sim$52GPa.nm (biaxial) independent of the
R0 or R45 conformation.


\begin{figure}[htb!]
 \centering
 \includegraphics[scale=0.35]{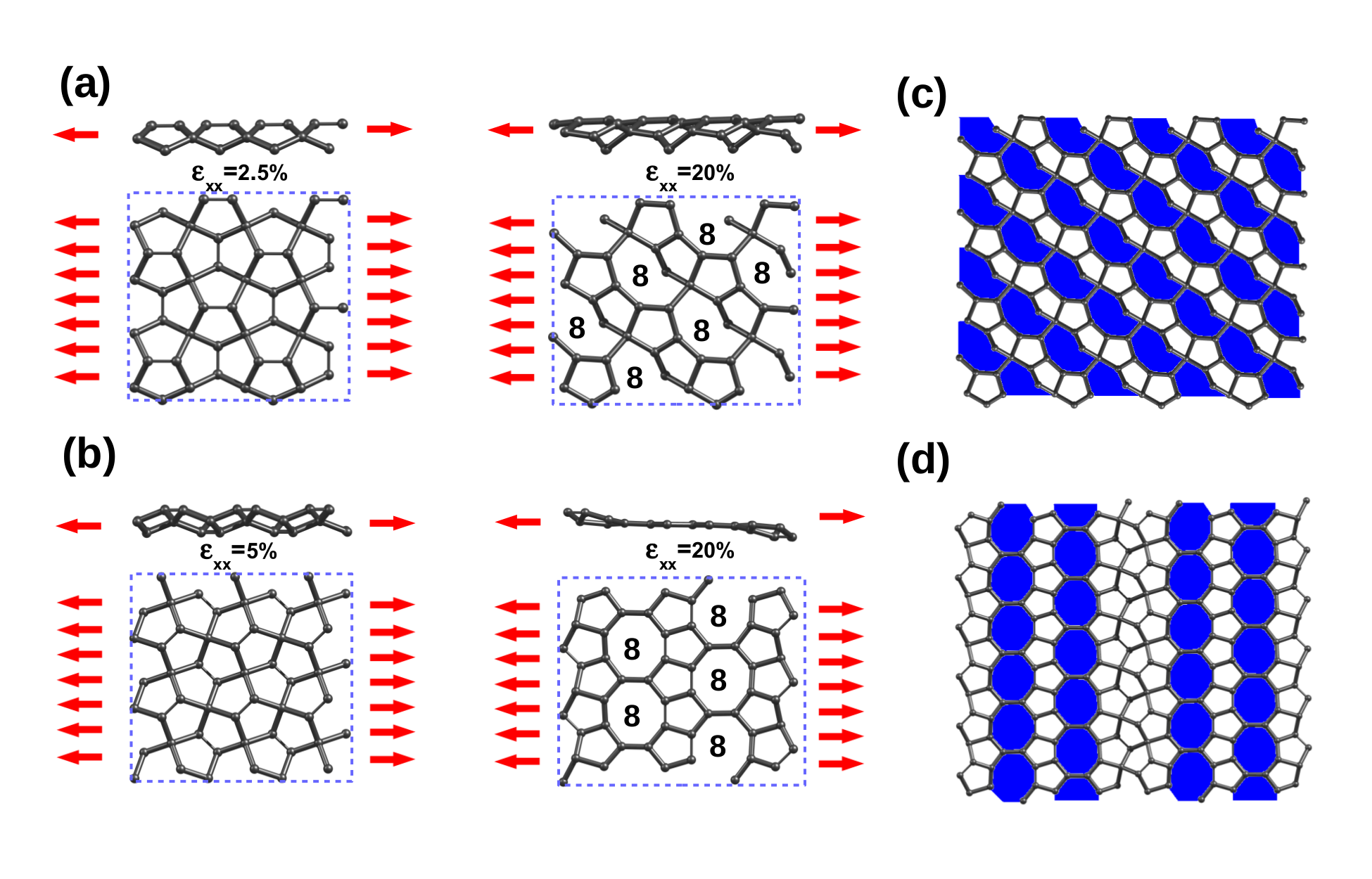}
 \caption{Snapshots of pentagraphene (PG) under 2.5\% and 20\% of uniaxial strain indicated by red arrows. 
   (a) R0 structure after uniaxial stretching converges to $8$-porous
   structures aligned to diagonal direction. A residual $\sigma_{xy}$ stress component
   is obtained which indicated that those structure after the failure point is not energetically stable.
   (b) R45 structure after uniaxial stretching converges to $8$-porous
   structures aligned to perpendicular direction. No residual $\sigma_{xy}$ stress component
   are observed which indicated that those structure after failure point is energetically stable.
   Extended PG porous membranes after failure
   strain are also shown. Uniaxial strain of $20\%$ along the
   $x$-axis for R0 structure (c) and R45 structure (d). Porous of $8$ (blue)
   carbon atoms rings are highlighted for better visualization. }
 \label{fig07}
\end{figure}

\begin{figure}[htb!]
 \centering
 \includegraphics[scale=0.45]{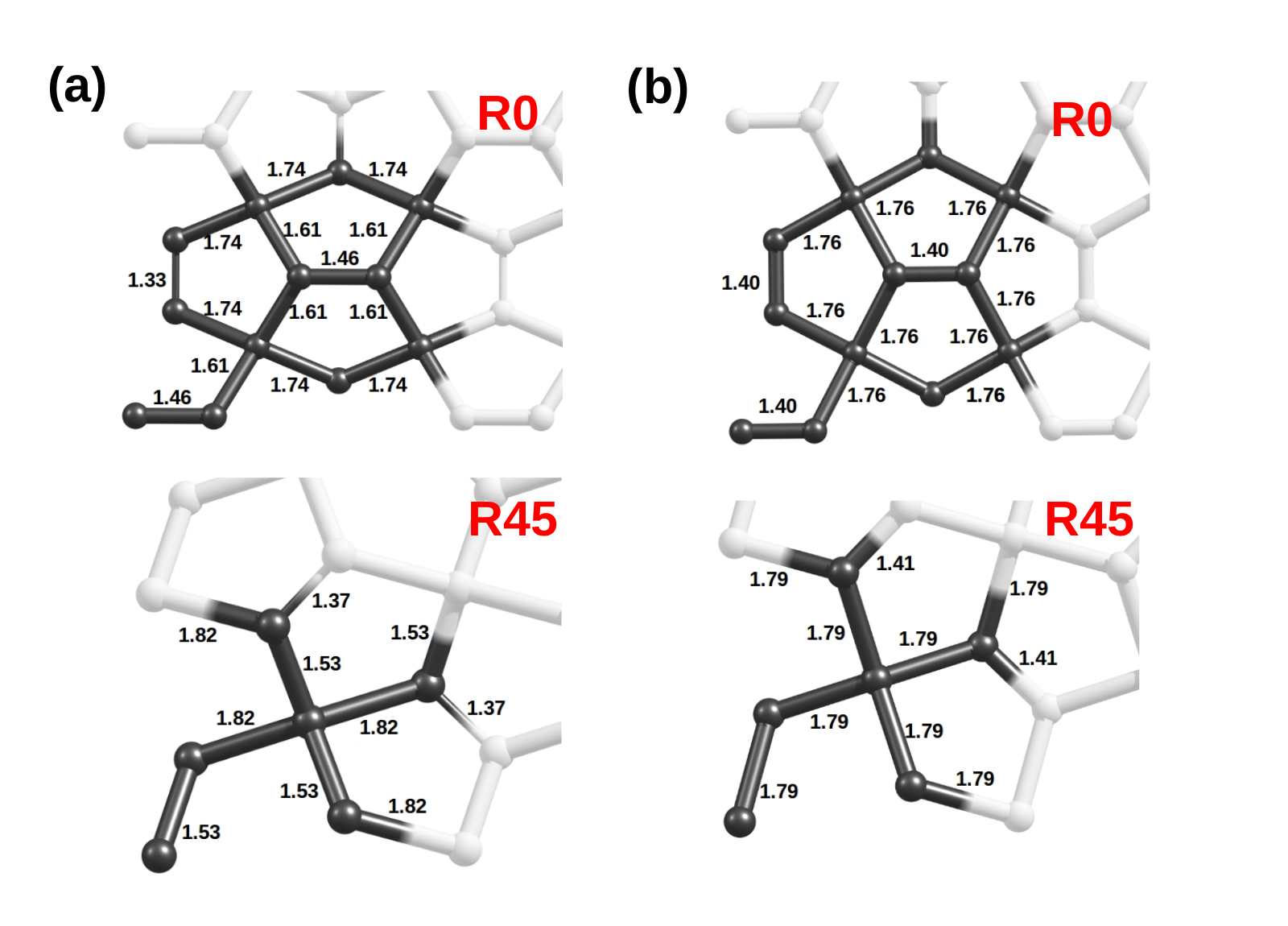}
 \caption{(a) R0 (top) and R45 (down) structures under 19\% of uniaxial strain. (b) R0 (top) and R45 (down) structures under 19\% of biaxial strain. Carbon-Carbon bonds lengths in \AA~units are shown in detail. Carbon atoms from primitive unit cell are highlighted in black
  for better visualization.}
 \label{fig08}
\end{figure}

In Fig.~\ref{fig07}a-b, we present PG snapshots for 
R0 and R45 structures, respectively before and after failure caused by a maximum of $20\%$ for $x$-axis
stretching (similar results were obtained for y-axis stretching, not shown). PG fractured membranes of Fig.~\ref{fig07}a-b
were duplicated along $x$- and $y$-direction in Fig~\ref{fig07}c-d for a better visualization of the fracture patterns 
(see colored non-pentagonal rings).
It is interesting to note that the patterns of R0 and R45 ruptured
structures after $20\%$ of strain are similar with the formation of
porous membranes with $8-C$ rings for both strained directions
(cf. Fig.~\ref{fig07}c-d).  However, we note significant differences
in R0 and R45 structures concerning the maximum of tensile stress
under uniaxial stretching just before the rupture, as we can see
in Fig.~\ref{fig06}c. We obtained $\sim$38GPa.nm (R0), and
$\sim$29GPa.nm (R45) for this quantity. We can explain such a
difference in terms of the parallel C-C bonds present in R0 structure,
which play a significant role at 19\% of strain, as one can see in
Fig.~\ref{fig08}a.

For symmetry reasons, the original PG structure (no stressed) has
only two types of C-C bonds, which connect both tri-coordinated atoms (1.33\AA)
and also tri- and tetra-coordinated atoms (1.55\AA), independent of the
R0 and R45 unit cell construction. However, when subjected
to uniaxial loading, those bonds are stressed differently.
At 19\% of strain, the parallel aligned C-C bonds in R0 structure are stressed up to 1.46\AA,
while the perpendicular C-C bonds remains with the same length of zero-strain structure,
as we can observe in Fig.~\ref{fig08}a (top panels). 
In R45 conformation, at 19\% of strain, the same C-C bonds are slightly elongated to 1.37\AA.
Therefore, we suggest that the triple coordinated carbon atoms connection in R0 structure
are directed affected by uniaxial strain and this conformation 
causes a larger stress obtained for R0 conformation when compared to R45 structures.
During the biaxial loading, as we can observe from Fig.~\ref{fig08}b, similiar
C-C bond lengths are obtained for tri- and tetra-coordinated atoms, which
can explain why stress-strain curves are very similar for R0 and R45 structure.


Our DFT results are closer to our MD simulations, rather than to those from reported in Ref.~\cite{cranford6to5}. 
Apart from differences related to temperature effects and sample size, the main reason for differences between the classical results lies in the softness of the material as predicted by the different simulation set parameters. Note that our MD prediction for $Y$ and that from Ref.~\cite{cranford6to5} are in opposite 
trends when compared to DFT. However, we observe no hexagons during the fracture 
process (as investigated by DFT), which does not suggest a structural transition 
from PG to hexagraphene. This allows us to argue that the Muller's set of ReaxFF  parameters is more suitable to describe the fracture of pentagraphene as its predictions are 
closer to what we expect from \textit{ab initio} calculations.


\newpage
\section{Summary and Conclusions}
\hspace*{0.50cm}In this work, we have investigated the mechanical properties and fracture patterns of a new carbon allotrope named pentagraphene (PG). We have combined DFT and reactive molecular dynamics simulations by using the well-known ReaxFF force field. Our results showed conflicting data depending on the set of parameters. DFT calculations help to explain these discrepancies and the Mueller's parameter set~\cite{mueller2010development} seems to provide more reliable results for this system.
Our results also help to provide further insigths on two conflicting literature issues regarding PG mechanical properties. The auxetic character (negative Poisson's ratio) was confirmed and the reported structural transition from PG to graphene ~\cite{cranford6to5} is not consistent with DFT results. 
We show that PG membranes can hold up to 20\% of strain and that fracture occurs
only after substantial dynamical bond breaking and the formation of 7, 8 and 11 carbon
rings and carbon chains were observed prior complete fracture. The stress-strain behavior was observed to follow two regimes, one exhibiting linear elasticity followed by a plastic one, this one involving carbon atom re-hybridization
with the formation of carbon rings and chains. Our MD results also show that mechanically
induced structural transitions from PG to graphene is unlikely to occur, in contrast to what
was previously predicted in the literature.

\newpage
\section{Acknowledgements}

This work was supported in part by the Brazilian Agencies CAPES, CNPq
and FAPESP. The authors thank the Center for Computational Engineering
and Sciences at Unicamp for financial support through the FAPESP/CEPID
Grant \#2013/08293-7. AFF is a fellow of the Brazilian Agency CNPq
(\#302750/2015-0) and acknowledges support from FAPESP grant
\#2016/00023-9. ECG acknowledges support from Conselho Nacional de
Desenvolvimento Cient\'ifico e Tecnol\'ogico (CNPq) (Process Number
473714/2013-2). ECG and JMS acknowledge support from Coordena\c c\~ao
de Aperfei\c coamento de Pessoal de N\'ivel Superior (CAPES) through
the Science Without Borders program (Project Number A085/2013). AGSF and JMS acknowledge 
PNPD CAPES fellowship.

\newpage
\section{References}

\bibliographystyle{elsarticle-num}

\end{document}